\documentclass{report}

\everydisplay{\rm }

\begin{document}
\begin{center}
\LARGE The Volume Operator  \\ [.125in] for Singly Polarized Gravity Waves \\[.125in]
with Planar or Cylindrical Symmetry \\ [.25in] \large Donald E.
Neville \footnote{\large Electronic address: dneville@temple.edu}
\\Department of Physics
\\Temple University
\\Philadelphia 19122, Pa. \\ [.25in]
April 26, 2006 \\ [.25in]
\end{center}

\newcommand{\E}[2]{\mbox{$\tilde{{\rm E}} ^{#1}_{#2}$}}
\newcommand{\A}[2]{\mbox{${\rm A}^{#1}_{#2}$}}
\newcommand{\Np}{\mbox{${\rm N}'$}}
\newcommand{\Etwo}{\mbox{$^{(2)}\!\tilde{\rm E} $}\ }
\newcommand{\Etld }{\mbox{$\tilde{\rm E}  $}\ }

\newcommand{\bea}{\begin{eqnarray}}
\newcommand{\eea}{\end{eqnarray}}
\newcommand{\be}{\begin{equation}}
\newcommand{\ee}{\end{equation}}
\newcommand{\nn}{\nonumber \\}
\newcommand{\rta}{\mbox{$\rightarrow$}}
\newcommand{\rla}{\mbox{$\leftrightarrow$}}
\newcommand{\eq}[1]{eq.~(\ref{#1})}
\newcommand{\Eq}[1]{Eq.~(\ref{#1})}
\newcommand{\eqs}[2]{eqs.~(\ref{#1}) and (\ref{#2})}

\large
\begin{center}
{\bf Abstract}
\end{center}
    A previous paper constructed a kinematic basis for spin networks
with planar or cylindrical symmetry
 and arbitrary polarization.
This paper imposes a constraint which limits the gravitational
wave to a single polarization.  The spectrum of the constraint
contains a physically reasonable number of zero eigenvalues, and
the zero eigenvectors can be constructed explicitly. Commutation
of the constraint with the Hamiltonian is expected to lead to a
further constraint. This new constraint is not investigated in
this paper, but I argue it will be non-local, relating states at
two or more neighboring vertices.
 \\[.125in]
PACS categories: 04.60.Pp, 04.30.-w

\section{Introduction}

    In a previous paper, I constructed a kinematic basis for a
spin network with planar or cylindrical symmetry and studied the
eigenfunctions and eigenvalues of the volume
operator\cite{paper1}. In that paper, the gravitational wave could
have either polarization. In this paper I study the volume
operator for a wave limited to a single polarization.

    In classical gravity, and in quantum gravity based on local
field theory, the assumption of a single polarization greatly
simplifies the discussion.  If both polarizations are present the
matrix of metric components (or triad components, if one is using
connection triad variables) has a 1x1 subblock and a 2x2 subblock
after gauge fixing.  The 1x1 subblock contains the longitudinal
component of the metric or triad (r component in the cylindrical
case; z component in the planar case); the 2x2 subblock contains
components transverse to the direction of the wave. In the single
polarization case, the matrix simplifies further, becoming
diagonal.  In terms of ADM metric connection variables, the
off-diagonal element $g_{xy}$ (or $g_{\theta\phi}) $=0.  In the
connection triad variables used in this paper, the equivalent
statement is \be
    Tr( \Etld^{x} \Etld^{y}) = 0, \label{Econstraint}
\ee or the analogous statement for cylindrical coordinates.

    In local field theory, when the constraint \eq{Econstraint}
is commuted with the Hamiltonian, one obtains a further constraint
which states that the conjugate ADM momentum $\pi^{xy}$ also
vanishes. In triad connection language, \bea
    0&=&Tr( \Etld^{x} \Etld^{a})Tr( \Etld^{y} A_{a}))+
            (x\leftrightarrow y)-\Gamma^{xy};\nonumber \\
    \Gamma^{xy} &=& -i Tr((\Etld^x \stackrel{\leftrightarrow}{\partial_z}
                    \Etld^y )\Etld^z).
                            \label{Aconstraint}
\eea The $\Gamma$ term subtracts out the spin connection part of
A. Imposing these constraints causes a degree of freedom and its
conjugate momentum to disappear entirely from the calculation,
which simplifies considerably.

    In the spin network case, it is straightforward to impose
\eq{Econstraint}.  This constraint resembles the volume operator,
in that it is constructed entirely from triads and is derivative
free. If one makes the simplest assumption that only the field
strengths (or other operators with derivatives) need to be
non-local, then one can take the operator of \eq{Econstraint} to
be local. That is, the operator acts at a single vertex, merely
reshuffling the holonomic basis states at that vertex.  The
constraint is no harder to handle than the volume operator, which
is also local. (All \Etld operators should be integrated over an
area, but this is straightforward and I suppress the area
integrations.)

    However, in a spin network approach, the constraint \eq{Aconstraint}
is unacceptable.  Connections A are not diffeomorphism invariant
unless integrated over an edge and included in a holonomy.  It is
also much easier to construct a field theory based on holonomies
than one based on A's, because holonomies may be described by a
compact manifold, the Euler angles \cite{Thiemann}.

    Perhaps one could propose a spin network generalization of
\eq{Aconstraint} that is local. (Promote the A operator to a
holonomy which acts at the same vertex as the \Etld operators in
\eq{Aconstraint}) However, I will give two arguments that the
holonomy (and perhaps some of the \Etld as well) will have to be
non-local.  That is, the holonomy which regularizes the A in
\eq{Aconstraint} will include segments joining two or more
neighboring vertices.

    (First argument.) Since the constraint \eq{Aconstraint} is a
momentum, the time derivative of the original constraint, the
natural way to obtain (the spin network version of)
\eq{Aconstraint} is to commute \eq{Econstraint} with the spin
network Hamiltonian. It is not clear at present what to use for
that Hamiltonian; but it must be non-local, in order for
propagation to occur. Commuting even a local object with a
non-local object is very likely to lead to a non local object.

    (Second argument). Consider the Killing
vectors which define planar symmetry.  (I am concentrating on the
planar case. The discussion for the cylindrical case is virtually
identical \cite{Kuc}.)  The space possesses two spacelike Killing
vectors which commute, and one can choose coordinates so that
these two vectors become $\partial/\partial x$ and
$\partial/\partial y$ . This planar symmetry by itself does not
imply one polarization only. Given only the existence of the two
Killing vectors, all three transverse components of the metric,
$g_{xx}, g_{yy}, g_{xy}$ or all four transverse triads can be
non-zero; this is enough to permit two polarizations.

    In classical general relativity, and in local quantum field theory
treatments of gravity, one must demand hypersurface orthogonality
of the Killing vectors in order to get one polarization .
Hypersurface orthogonality can be shown to imply that $g_{xy}$ and
its canonical momentum vanish \cite{Kuc}.

    Hypersurface orthogonality is an intrinsically non-local
idea.  If  a Killing vector field such as $\partial/\partial x$
exists, then in each plane z = constant = $z_1$ there exist
one-dimensional integral curves with the Killing vector as tangent
vector. The plane at z = $z_2$ possesses similar integral curves,
but no relation is implied between the two sets of curves at $z_1$
and $z_2$. The additional assumption of hypersurface orthogonality
creates such a relation. The two sets of integral curves must be
normal to a common set of hyperplanes . The integral curves can be
described as curves of increasing x, and the hyperplanes as
hyperplanes of constant x.

    Without hypersurface orthogonality, the integral curves of
$\partial/\partial x$ may twist in any manner as one goes from one
value of z to the next. With hypersurface orthogonality, this
twisting is strongly constrained.

    In the spin network context, going from one value of z to the
next is equivalent to going from one vertex to the next.
Hypersurface orthogonality implies a relation between structures
at two different vertices, necessarily a non-local relation. (End
of second argument).

    If \eq{Aconstraint} must be promoted to a non-local constraint,
then within a spin network context,  the one-polarization case is
harder than the two-polarization case. In this paper I carry the
discussion of the one-polarization case as far as I can, given
that I have a kinematic basis but no Hamiltonian, therefore no way
to construct \eq{Aconstraint}.  I impose \eq{Econstraint} on the
states of the kinematic basis, but not \eq{Aconstraint}.  I then
ask which states of the two-polarization case are ruled out by the
constraint.

    For orientation, it is helpful
to keep in mind an example from quantum mechanics, the two
dimensional simple harmonic oscillator with Hamiltonian \be
    H = -(\hbar^2/2m)[\partial^2/\partial x^2 +\partial^2/\partial
    y^2]
        + (k/2)(x^2 + y^2).
\ee This problem is separable, with eigenfunctions $\mid n_x n_y>$
labelled by occupation numbers.  Suppose I wish to impose the
constraint that the oscillator possesses only a single
"polarization", say it oscillates only along x.  I can impose the
constraint on the classical theory first, then quantize
(constraint first). Alternatively, I can follow Dirac and quantize
first, then impose the constraint (quantize first) \cite{Dirac}.
For reasons to be explored in the next section, I will choose to
follow the second, Dirac procedure and quantize first.

    Suppose I apply the Dirac procedure to the oscillator problem.
I know the answer I want: the constraints should rule out all
states except those of the form $ \mid n_x;n_y = 0>$. However, if
I require physical states to obey  \be
    y \mid > = p_y \mid > = 0,\label{strongy}
\ee I find these requirements are too strong.  From the theory of
coherent states,  there is no ket which satisfies even one of
these equations \cite{Glauber}. (Recall that both coordinate and
momentum  contain creation operators, which make it difficult for
these operators to completely destroy any state.)

    I can try the milder constraint,
\begin{equation}
     <\mid y \mid>  = 0,\label{avgy}
\end{equation}
 which is perhaps enough to guarantee a satisfactory
classical limit.  However, whereas the previous constraint was too
strong, this constraint is too weak. For example, it is satisfied
by any state $\mid >$ which has definite parity,  including states
with $n_y$ non-zero. To get the answer I want, I must impose at
least \eq{avgy} and \be
    <\mid y^2\mid > = 0.  \label{avgy2}
\ee \Eq{avgy2} gives the energy  a satisfactory classical limit. I
conclude a straightforward application of the constraint to a ket,
the analog of \eq{strongy}, may not work; I may have to impose
averaged constraints, \eqs{avgy}{avgy2}.

    For another example, this time from field theory, consider
quantization of the electromagnetic field in Lorentz gauge.  One
must impose both of the constraints \bea
    0&=& <\mid \partial \cdot A \mid>; \nonumber \\
    0&=& <\mid (\partial \cdot A)^2 \mid>,\label{Lorentz}
\eea in order to get the correct classical limit, including the
correct classical energy \cite{Heitler}. Again, \be
    0 =  \partial \cdot A \mid >
\ee is too strong.

    These examples determine much of the discussion in section two.
First I discuss the relative merits of constraint first vs.
quantize first approaches, and  I opt for the latter. I then
impose the analog of \eq{strongy}, \be
    Tr( \Etld^{x} \Etld^{y})\mid> = 0.\label{strongE}
\ee In the oscillator and Lorentz gauge examples, constraints of
this type have no solutions. However, in the spin network case I
was able to find normalizable states satisfying \eq{strongE} . The
kernel of the constraint is non-trivial. As a check, I impose the
constraints in the average sense: \bea
    0&=&<\mid \mbox{Tr}( \Etld^{x} \Etld^{y})\mid> \nonumber \\
    &=&<\mid [\mbox{Tr}( \Etld^{x} \Etld^{y})]^2\mid>. \label{avgE}
\eea I find that the states which satisfy these constraints are
the same as the states which satisfy \eq{strongE}.

\section{Dirac Quantization of the One-Polarization Case}

    I can impose the constraint on the classical theory first, then
quantize the single-polarization classical theory (constrain
first); or I can quantize first, then impose the constraint as a
condition on the quantized states (quantize first; Dirac
quantization). If I constrain first, it is natural to follow
Bojowald and introduce fields which have the constraint built in
\cite{BojBH}. \bea
    \E{x}{A}S_A &=& \tilde{E}^x[\cos(\alpha+\beta)S_1 +
        \sin(\alpha+\beta)S_2];\nonumber \\
    \E{y}{A}S_A &=& \tilde{E}^y[-\sin(\alpha+\beta)S_1 +
        \cos(\alpha+\beta)S_2];\nonumber \\
    \A{A}{x}S_A&=& A_x [\cos(\beta)S_1 + \sin(\beta)S_2];
                        \nonumber \\
    \A{A}{y}S_A&=& A_y [-\sin(\beta)S_1 + \cos(\beta)S_2].
                        \label{polar}
\eea The first two lines, for example, relate the four "Cartesian"
fields   (\E{x}{A} and \E{y}{A}) to  three "polar" fields (the
magnitudes $E^x,E^y,$ and the angle $\alpha+\beta$).  One degree
of freedom is lost, but this is acceptable: the new polar fields
have the constraint Tr$( \Etld^{x} \Etld^{y})=0$ built in. From
\eq{polar} the A's obey a similar constraint, \be
    \mbox{Tr} A_x A_y  = 0, \label{polAconstr}
\ee which again removes one degree of freedom, and is easier to
implement than the constraint of \eq{Aconstraint}.  Bojowald
carries out the canonical transformation from Cartesian to polar
fields and finds the following pairs of canonical variables
\cite{BojBH}. \bea
    && (E^x \cos(\alpha)= P^x, A_x);\nn
    &&  (E^y \cos(\alpha)= P^y, A_y);\nn
    && (\sin \alpha (A_xP^x+A_yP^y) = P^{\beta},
    \beta);\nn
    && (\E{z}{Z},\A{Z}{z}).\label{canpairs}
\eea  $P^{\beta}$ turns out to be  the Gauss Constraint.  The
angle $\alpha$ is not an independent degree of freedom, but is
given by
\[
    \tan \alpha = P^{\beta}/(A_xP^x+A_yP^y). \]
These are classical field theory definitions.  It is not clear how
to order the operators in $\tan \alpha$ after quantizing, or what
the spin network definitions should be.

    The lack of precise definitions is already a problem at the
kinematic level, since the formula for the volume operator in
polar coordinates involves $\alpha$. \bea
    V^2 &=& \E{z}{Z} E^x E^y \nonumber \\
        &=& \E{z}{Z} P^x P^y /(cos \alpha)^2. \label{polvol}
\eea Since I was unable to supply any convincing spin network
definitions of $\alpha$ or $V^2$, I determined to avoid polar
coordinates as much as possible.  Every time I attempted a
"constrain first" quantization, however, I found myself
introducing the polar variables, explicitly or implicitly.  A
constrain first quantization which avoids polar fields may be
possible; but I was unable to find it.

    For the rest of this paper I will use the  quantize
first approach.  This  has the advantage that the first steps
(gauge fixing, quantization, kinematic basis and dot product) have
been worked out already in the previous paper; and the polar
fields need not be used at any stage.

    I now impose the constraint, in its unaveraged form, \eq{strongE}.
I take the ket to be a linear combination of states in the
kinematic basis. Note that the trace in \eq{strongE}, is very
similar in structure to the 2x2 transverse subblock of the (square
of the) volume operator: \bea
    (V_2)^2 &=& \epsilon_{ZAB}\E{x}{A}\E{y}{B}\nonumber \\
            &=& i(\E{x}{+}\E{y}{-} -\E{x}{-}\E{y}{+});\nonumber \\
    E^{xy} &:=& 2 Tr( \Etld^x \Etld^y) \nonumber \\
            &=& \E{x}{+}\E{y}{-} + \E{x}{-}\E{y}{+}.
            \label{V2vsExy}
\eea on the second lines I have used the linear combinations with
simple U(1) transformation properties,\[\E{x}{\pm}= (\E{x}{X}\pm
i\E{x}{Y})/\sqrt{2}.\]  The two expansions in \eq{V2vsExy} are
identical, except for the extra i's and the extra minus sign in
$(V_2)^2$. Therefore I can get the action of the constraint from a
simple modification of the action of the $(V_2)^2$ operator.

    First recall the equation for the action of $(V_2)^2$.  Write
an eigenfunction of this operator as a linear combination of
states in the kinematic basis: \be
     \mid \vec{c}(\lambda;L_x,L_y,F)>=\Sigma_D
     c(D;\lambda,L_x,L_y,F)Y_{L_x m_x}Y_{L_y m_y}.\label{Yexpand}
\ee $\lambda$ is the eigenvalue, $\vec{c}$ is the eigenvector with
components \{c(D;$\lambda,L_x,L_y$,F)\}, the Y's are the spherical
harmonics constructed in \cite{paper1}, and \bea
    F&=&(m_x+m_y)/2 ; \nonumber \\
    D&=& (m_x-m_y)/2. \label{DFdef}
\eea The eigenfunction contains a sum over D only; the volume
operator does not change F. From the previous paper, $(V_2)^2$
acting on this state leads to the following recurrence relation
for the c's. \bea
    2\lambda ~c(D;\lambda) &=& i g_- ~c(D-1;\lambda) - i g_+ ~c(D+1;\lambda);\nonumber \\
       g_-(D) &=& \sqrt{(L_x-F-D+1)(L_x+ F + D)}\cdot \nonumber \\
         & & \sqrt{(L_y+F-D+1)(L_y-F+D)}; \nonumber \\
         g_+(D)&=& g_-(D+1). \label{V2def}
\eea (For simplicity I have suppressed some of the arguments of
the c's.)  To obtain the action of $E^{xy}$, I use \eq{V2vsExy}. I
drop the i's and the minus sign in \eq{V2def}.  Also, I replace
the eigenvalue $\lambda$ by zero, since $E^{xy} \psi = 0$.\be
    0 = g_- e(D-1) + g_+ e(D+1).\label{Exydef}
\ee I denote the expansion coefficients by e, rather than c,
because the c's (the volume eigenfunctions) in general will not
satisfy \eq{Exydef}.

    The following lemma is straightforward to prove.

Lemma . If the eigenvector $\vec{c}(\lambda$) with components
\{c(D;$\lambda$)\} is a solution to \eq{V2def}, then the
eigenvectors \be
    \vec{e}(\pm \lambda) = \{(\mp i)^D c(D;\lambda)\} \label{edef}
\ee are eigenvectors of \eq{Exydef} with eigenvalues $\pm
\lambda$, i.e. \be
    \pm 2\lambda e(D;\pm \lambda) = g_- e(D-1;\pm \lambda) + g_+
e(D+1;\pm \lambda). \label{epmeqn} \ee

    This lemma implies that $(V_2)^2$ and the
constraint operator have exactly the same eigenvalue spectrum
(although their eigenvectors are not the same). (For example, in
the previous paper I proved that the non-zero eigenvalues of
$(V_2)^2$ occur in pairs $(\lambda,-\lambda)$; and the lemma
indicates a similar pairing for the eigenvalues of the constraint
operator.)

    From the previous paper, $(V_2)^2$ has zero eigenvalues \cite{paper1}. Hence the
constraint operator has zero eigenvalues, and the constraint
equation \eq{strongE} has non-trivial solutions.

    From the discussion in the introduction, I should also investigate
the averaged constraints, \eq{avgE}.  It is easy to see that there
are a large number of solutions to the first, linear constraint:
the combinations $\mid \vec{e}(\lambda)> \pm \mid
\vec{e}(-\lambda) >$ are sent into the orthogonal combinations
$\mid \vec{e}(\lambda)> \mp \mid \vec{e}(-\lambda) >$ by the
constraint. (As at \eq{Yexpand}, $\mid \vec{e}(\lambda)>$ is
shorthand for the eigenstate of the constraint operator $\Sigma_D
e(D;\lambda) Y_{L_x m_x}Y_{L_y m_y}$.)

    However, the second, quadratic constraint in \eq{avgE}
is more restrictive. Let the quadratic constraint act on an
arbitrary state
\[
    \mid > = \sum_{\lambda} a(\lambda) \mid \vec{e}(\lambda)>.\]
The expectation value of the quadratic constraint is then\[
    \Sigma_{\lambda}\lambda^2 \mid a(\lambda)\mid^2.\]
This is not zero unless all of the $\lambda$ are zero. The
averaged constraints, \eq{avgE}, give the same result as the
unaveraged constraint, \eq{strongE}.

    In the oscillator example discussed in the introduction, the
imposition of a constraint lowers the dimension of the Hilbert
space signifigantly.  The unconstrained oscillator states are
labeled by two integers ($n_x,n_y$), and there are
($\mbox{countable infinity})^2$  states in the Hilbert space.
After the constraint restricts the states to $n_y$ = 0 only, the
number of admissible states drops to  ($\mbox{countable
infinity})^1$.  Roughly speaking, this is the square root of the
original number of states.

    Something similar happens in the present case.  Consider the
set of kinematic basis states \{$Y_{L_x m_x}Y_{L_y m_y}; \forall
m_x,m_y$\}.  Before imposition of the constraint, there are \[
    (2L_x +1)(2L_y+1)  ~\mbox{unconstrained states}\]
in this set.  After imposition of the constraint, the set contains
only those linear combinations which are eigenstates of the
constraint with eigenvector zero.  From the lemma, the volume and
constraint operators have exactly the same number of zero
eigenvalues.  The necessary condition for the volume operator to
have a zero was worked out in the previous paper \cite{paper1}.
The eigenvectors are sums of the form \eq{Yexpand}: sums over D
with F held fixed. There will be a zero eigenvector iff the number
of D's in the sum is odd. Therefore I can compute the number of
zero eigenvectors by constructing the following rectangular
lattice of points. Make the x-axis $m_x$ and y-axis $m_y$; draw
one point with coordinates ($m_x,m_y$), for each allowed pair of
m's, $(2L_x +1)(2L_y+1)$ points in all.

    In this lattice, diagonals at 135 degrees are lines of
constant F = $(m_x+m_y)/2$, while diagonals at 45 degrees are
lines of constant D = $(m_x-m_y)/2$.  The number of zeros is the
number of 135 degree diagonals which contain an odd number of
points (corresponding to an odd number of D's in the sum over D
with F held fixed).  After some experimenting with specific
examples, one arrives at the general formula for the number of
diagonals with an odd number of points = number of states
surviving the constraint.\[
    2~ max (L_x,L_y) + 1 ~\mbox{constrained states}\]
We have gone from order $(2L)^2$ states to order $(2L)^1$ states.
This is roughly a square root, similar to the oscillator example
and therefore not unreasonable.

    In the previous paper I computed the zero eigenvectors of the
volume operator \cite{paper1}.  Given that result, and the lemma,
it is possible to compute the zero eigenvectors of the constraint.
From the previous paper, the zero eigenvectors of the volume
operator have components\bea
    c(D;\lambda =0) &= & N \sqrt{f(D-1)/f(D)};\nonumber \\
    f(D)&=&
    \left(\frac{L_y-F+D}{2}\right)!\left(\frac{L_y+F-D}{2}\right)!\times\nonumber
    \\
    & &
    \left(\frac{L_x-F-D}{2}\right)!\left(\frac{L_x+F+D}{2}\right)!,\label{0eig}
\eea for D = max D, max D - 2, max D - 4, $\ldots$, min D; and
c(D) = 0  otherwise.  From the lemma, \eq{edef}, the components of
the corresponding zero eigenvector of the constraint are just \[
    e(D;\lambda = 0) = (i)^D c(D;\lambda = 0).\]
(The two states e($\pm;\lambda = 0)$ in \eq{edef} give the same
zero eigenvector, except perhaps for an overall minus sign,
because c(D;$\lambda $= 0) is zero for every other value of D.)

    These zero eigenvectors illustrate a point
made earlier, that volume and constraint operators have (the same
eigenvalues but) very different eigenvectors.  Zero constraint
eigenvalue does not imply zero volume eigenvalue.  An eigenvector
of the constraint operator with zero eigenvalue is a linear
combination of eigenvectors of the volume operator, with
eigenvalues which are in general different from zero.  Although
the volume operator and constraint have the same eigenvalue
spectrum, they do not commute.

    I return to the point made in the introduction: in classical
or quantum field theory, the single polarization case is simpler;
whereas for spin networks, the single polarization case is harder.
For example, experience with field theory might suggest using
eigenfunctions of the constraint as a basis; however, these
eigenfunctions are not eigenfunctions of the volume, and the spin
network Hamiltonian would be hard to evaluate in that basis.

    There is a further complication.  The constraint I have been
studying in this paper is a local constraint.  The Hamiltonian
acting on an eigenfunction of this local constraint will produce
another local constraint eigenfunction only if the state satisfies
the spin network version of  \eq{Aconstraint}.  As argued in the
introduction, that constraint is likely to be non-local.  Given
the problems with choosing a basis and the existence of a
non-local constraint, it would not be surprising if solutions for
a general polarization were obtained first, before solutions for a
single polarization.

    There may be ways around these complications, however.  In the
classical limit, the spin vectors associated with the $A_x$ and
$A_y$ holonomies will have sharp expectation values with minimal
uncertainties.  The local constraint implies that the two spin
vectors are perpendicular; while the volume eigenvalue is given by
the cross product of the two spin vectors.  It is not possible for
the classical state to be simultaneously an exact eigenfunction of
both the volume operator and the local constraint; but in the
classical limit it should be possible to find states which are
eigenfunctions of both constraints to very good approximation.
Further, the non-local constraint, when combined with the
Hamiltonian, may simplify the Hamiltonian. It is too early to
tell.  Much work remains to be done.

\end{document}